\documentclass{article}
\begin{document}
\title{Correlation-Based And-Operations Can Be Copulas: A Proof}
\author{Enrique Miralles-Dolz$^{1,2}$, Ander Gray$^{1,2}$, Edoardo Patelli$^3$,\\
Scott Ferson$^2$, Vladik Kreinovich$^4$, and Olga Kosheleva$^5$\\
$^1$Institute for Risk and Uncertainty, University of Liverpool,\\Liverpool, UK,
\{enmidol,akgray,ferson\}@liverpool.ac.uk\\
$^2$United Kingdom Atomic Energy Authority, Abingdon, UK\\
$^3$Centre for Intelligent Infrastructure, University of Strathclyde,\\Glasgow, UK,
edoardo.patelli@strath.ac.uk\\
$^4$Department of Computer Science, University of Texas at El Paso,\\El Paso, Texas 79968, USA, vladik@utep.edu\\
$^5$Department of Teacher Education, University of Texas at El Paso,\\El Paso, Texas 79968, USA, olgak@utep.edu\\
}
\date{}
\maketitle

\begin{abstract}
In many practical situations, we know the probabilities $a$ and $b$ of two events $A$ and $B$, and we want to estimate the joint probability ${\rm Prob}(A\,\&\,B)$. The algorithm that estimates the joint probability based on the known values $a$ and $b$ is called an and-operation. An important case when such a reconstruction is possible is when we know the correlation between $A$ and $B$; we call the resulting and-operation correlation-based. On the other hand, in statistics, there is a widely used class of and-operations known as copulas. Empirical evidence seems to indicate that the correlation-based and-operation derived in \cite{Miralles 2022} is a copula, but until now, no proof of this statement was available. In this paper, we provide such a proof.
\end{abstract}

\section{Formulation of the problem}

\noindent{\bf Correlation-based ``and"-operation.} In many practical situations, we know the probabilities $a$ and $b$ of two events $A$ and $B$, and we need to estimate the joint probability ${\rm Prob}(A\,\&\,B)$. An algorithm $f_\&(a,b)$ that transforms the known values $a$ and $b$ into such an estimate is usually called an {\it and-operation}.

One important case when such an estimate is possible is when, in addition to the probabilities $a$ and $b$, we also know the correlation $\rho$ between the corresponding two random events. It is known (see, e.g., \cite{Lucas 1995,Miralles 2022}) that in this case, we can uniquely determine the probability of ${\rm Prob}(A\,\&\,B)$ as
$$a\cdot b+\rho\cdot \sqrt{a\cdot (1-a)\cdot b\cdot (1-b)}.\eqno{(1)}$$

While this formula is true whenever the correlation is known, this formula does not lead to an everywhere defined and-operation. For example, for $a=b=0.1$ and $\rho=-1$, this formula leads to a meaningless negative probability
$$0.1\cdot 0.1+(-1)\cdot \sqrt{0.1\cdot 0.9\cdot 0.1\cdot 0.9}=0.01-0.09=-0.08<0.$$
To avoid such meaningless estimates, we need to take into account that the joint probability ${\rm Prob}(A\,\&\,B)$ must satisfy Fr\'echet inequalities (see, e.g., \cite{Frechet 1935}):
$$\max(a+b-1,0)\le {\rm Prob}(A\,\&\,B)\le\min(a,b).\eqno{(2)}$$
So, if an expert claims to know the correlation $\rho$ and the estimate for ${\rm Prob}(A\,\&\,B)$ based on this value $\rho$ is smaller than the lower bound $\max(a+b-1,0)$ -- which cannot be -- a reasonable idea is to take the closest possible value of the joint probability, i.e., the value $\max(a+b-1,0)$. Similarly, if the estimate for ${\rm Prob}(A\,\&\,B)$ based on the expert-provided value $\rho$ is larger than the upper bound $\min(a,b)$ -- which also cannot be -- a reasonable idea is to take the closest possible value of the joint probability, i.e., the value $\min(a,b)$. Thus, we arrive at the following and-operation -- which we will call {\it correlation-based and-operation}:
$$f_\rho(a,b)=T_{a,b}\left(a\cdot b+\rho\cdot \sqrt{a\cdot (1-a)\cdot b\cdot (1-b)}\right),\eqno{(3)}$$
where
$$T_{a,b}(c)=\max(a+b-1,0)\mbox{  if  } c<\max(a+b-1,0);$$
$$T_{a,b}(c)=c \mbox{  if  } \max(a+b-1,0)\le c\le\min(a,b);\mbox{  and}\eqno{(4)}$$
$$T_{a,b}(c)=\min(a,b)\mbox{  if  } \min(a,b)<c.$$
\medskip

\noindent{\bf Question: is this and-operation a copula?} In probability theory, there is a known class of and-operations known as {\it copulas} (see, e.g., \cite{Nelsen 2007,Schweizer 2011}). These are functions $C(a,b)$ for which, for some random 2-D vector $(X,Y)$, the joint cumulative distribution function $F_{XY}(x,y)\stackrel{\rm def}{=}{\rm Prob}(X\le x\,\&\,Y\le y)$ has the form $F_{XY}(x,y)=C(F_X(x),F_Y(y))$, where $F_X(x)\stackrel{\rm def}{=}{\rm Prob}(X\le x)$ and $F_Y(y)\stackrel{\rm def}{=}{\rm Prob}(Y\le y)$ are known as {\it marginals}.

One important aspect of (3)-(4) is that these formulas can be expressed as a copula (2-copula) family as described in \cite{Miralles 2022}, allowing us to operate not only with precise probabilities, but also with interval probabilities and probability boxes.
A 2-copula must satisfy the following properties:
\begin{enumerate}
    \item Grounded: $C(0,b)=C(a,0)=0$
    \item Uniform margins: $C(a,1)=a;C(1,b)=b$
    \item 2-increasing: $C(\overline a,\overline b)+C(\underline a,\underline b)-C(\overline a,\underline b)-C(\underline a,\overline b)\ge 0$ for all $\underline a<\overline a$ and $\underline b<\overline b$
\end{enumerate}

It is easy to see that (3)-(4) satisfies the two first properties. In \cite{Miralles 2022} the third property was checked for a dense set of tuples $(\underline a,\overline a,\underline b,\overline b,\rho)$, and for all these tuples, the inequality was satisfied. However, at that moment, we could not prove that the correlation-based and-operation is indeed a 2-copula.
In this paper we provide the missing proof.

\section{Main result}

\noindent{\bf Proposition.} {\it For every $\rho\in[-1,1]$, the correlation and-operation $f_\rho(a,b)$ described by the formulas (3)-(4) is a copula.}
\medskip

\noindent{\bf Proof.}
\medskip

\noindent $1^\circ$. It is known that the desired inequality has the following property -- if we represent a box $[\underline a,\overline a]\times[\underline b,\overline b]$ as a union of several sub-boxes, then the left-hand side of the desired inequality is equal to the sum of the left-hand sides corresponding to sub-boxes. 

Indeed, as one can easily check, there is the following {\it additivity} property: for each box consisting of several sub-boxes, the left-hand side of the inequality (4a) that corresponds to the larger box is equal to the sum of expressions (4a) corresponding to sub-boxes. Thus, if the expressions corresponding to sub-boxes are non-negative, then the expression (4a) corresponding to the larger box is also non-negative.

In general, the and-operation described by the formula (4) has three different expressions. So, to prove that the expression (4a) corresponding to this expression is also non-negative, we need to consider cases when at different vertices of the box, we may have different expressions. Good news is that every box whose vertices are described by different expressions can be represented as the union of sub-boxes in which:
\begin{itemize}
\item either all vertices are described by the same expression 
\item or two vertices are on the boundary between the areas of different expressions. 
\end{itemize}
This is easy to see visually: the following box, in which the slanted line represents the boundary between the areas
\medskip

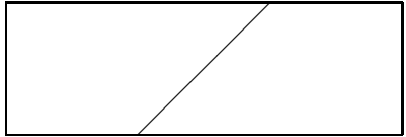
\begin{center}
\begin{picture}(150,50)
\put(0,0){\line(1,0){150}}
\put(0,50){\line(1,0){150}}
\put(0,0){\line(0,1){50}}
\put(150,0){\line(0,1){50}}
\put(50,0){\line(1,1){50}}
\end{picture}
\end{center}
\medskip

\noindent can be represented as the union of sub-boxes with the desired property:
\medskip

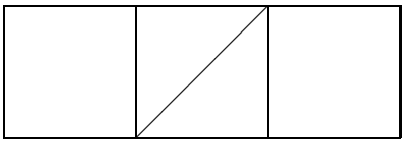
\begin{center}
\begin{picture}(150,50)
\put(0,0){\line(1,0){150}}
\put(0,50){\line(1,0){150}}
\put(0,0){\line(0,1){50}}
\put(150,0){\line(0,1){50}}
\put(50,0){\line(1,1){50}}
\put(50,0){\line(0,1){50}}
\put(100,0){\line(0,1){50}}
\end{picture}
\end{center}
\medskip

Thus, to prove that our and-operation is a copula, it is sufficient to consider only boxes of the following type:
\begin{itemize}
\item boxes for which all four vertices belong to the same area, and 
\item boxes for which two vertices belong to the boundary between two areas.
\end{itemize}
The functions $\max(a+b-1,0)$ and $\min(a,b)$ are known to be copulas, so if all four vertices belong to one of these areas, then the desired inequality (4a) is satisfied. So, it is sufficient to consider:
\begin{itemize}
\item boxes for which all four vertices belong to the new area, in which the and-operation is described by the expression (1); we will consider such boxes in Parts 2--4 of this proof, and 
\item boxes for which two vertices belong to the boundary between two areas; these boxes will be considered in the following Parts of the proof. 
\end{itemize}
\medskip

\noindent $2^\circ$. Let us start by considering boxes for which all four vertices belongs to the area in which the and-operation is described by the formula (1). 
\medskip

It is known \cite{Durante 2010} -- and it is easy to prove by considering infinitesimal differences $\overline x-\underline x$ and $\overline y-\underline y$ -- that for smooth functions, the desired inequality is equivalent to the fact that the partial derivative $$\frac{\partial C}{\partial a}$$ is non-decreasing in $b$, i.e., equivalently, that the mixed derivative is non-negative: $$d\stackrel{\rm def}{=}\frac{\partial^2 C}{\partial a\,\partial b}\ge 0.$$ Thus, to prove that $f_\rho(a,b)$ is a copula, it is sufficient to prove that its mixed derivative is non-negative everywhere where the new formula is applied.

Indeed, at the points where the formula (1) is applied, the derivative of $f_\rho(a,b)$ with respect to $a$ has the has the form
$$\frac{\partial f_\rho}{\partial a}=b+\rho\cdot \frac{1-2\cdot a}{2\cdot\sqrt{a\cdot (1-a)}}\cdot \sqrt{b\cdot (1-b)},\eqno{(4{\rm b})}$$ and thus, the mixed derivative has the following form:
$$d=\frac{\partial}{\partial b}\left(\frac{\partial f_\rho}{\partial a}\right)=1+\rho\cdot \frac{(1-2\cdot a)\cdot (1-2\cdot b)}
{4\cdot \sqrt{a\cdot (1-a)\cdot b\cdot (1-b)}}.\eqno{(5)}$$
Since the expression (1) does not change if we swap $a$ and $b$, it is sufficient to consider the case when $a\le b$.

When $\rho=0$, we get a known copula $f_0(a,b)=a\cdot b$. So, it is sufficient to consider cases when $\rho\ne 0$. This can happen
when $\rho>0$ and when $\rho<0$. Let us consider these cases one by one.
\medskip

\noindent $3^\circ$. Let us first consider the case when $\rho>0$. 
\medskip

In this case, since $a\le b$, we have $\min(a,b)=a$ and thus, the condition (4) takes the form
$$a\cdot b+\rho\cdot \sqrt{a\cdot (1-a)\cdot b\cdot (1-b)}\le a,\eqno{(6)}$$
i.e., equivalently,
$$\rho\cdot \sqrt{a\cdot (1-a)\cdot b\cdot (1-b)}\le a-a\cdot b=a\cdot (1-b)\eqno{(7)}$$
and thus,
$$\rho\le \frac{a\cdot (1-b)}{\sqrt{a\cdot (1-a)\cdot b\cdot (1-b)}}=\frac{\sqrt{a\cdot (1-b)}}{\sqrt{(1-a)\cdot b}}.\eqno{(8)}$$
For all such $\rho$, we need to prove that the expression (5) is non-negative.

When both $a$ and $b$ are larger than 0.5 or both are smaller than 0.5, the differences $1-2a$ and $1-2b$ have the same sign and thus, their product is non-negative and the expression (5) is non-negative. So, the only case when we need to check that $d\ge 0$ is when one of the two values $a$ and $b$ is smaller than 0.5 and another one is larger than 0.5. Since $a\le 0.5$, this means that $a<0.5<b$. In this case, the condition $d\ge 0$ takes the form
$$1-\rho\cdot \frac{(1-2\cdot a)\cdot (2\cdot b-1)}{4\cdot\sqrt{a\cdot (1-a)\cdot b\cdot (1-b)}}\ge 0,\eqno{(9)}$$
i.e., equivalently,
$$\rho\cdot \frac{(1-2\cdot a)\cdot (2\cdot b-1)}{4\cdot \sqrt{a\cdot (1-a)\cdot b\cdot (1-b)}}\le 1,\eqno{(10)}$$ and
$$\rho\le\frac{4\cdot \sqrt{a\cdot (1-a)\cdot b\cdot (1-b)}}{(1-2\cdot a)\cdot (2\cdot b-1)}.\eqno{(11)}$$
So, to prove that we always have $d\ge 0$, we need to prove that every $\rho$ that satisfies the inequality (8) also satisfies the inequality (11). Clearly, if some value $\rho$ satisfies the inequality (11), then every smaller value $\rho$ also satisfies this inequality. Thus, to prove the desired implication, it is sufficient to check that the inequality (11) is satisfied for the largest possible value $\rho$ that satisfies the inequality (8), i.e., for the value $\rho$ which is equal to the right-hand side of the inequality (8). For this $\rho$, the desired inequality (11) takes the form
$$\frac{\sqrt{a\cdot (1-b)}}{\sqrt{(1-a)\cdot b}}\le \frac{4\cdot \sqrt{a\cdot (1-a)\cdot b\cdot (1-b)}}{(1-2\cdot a)\cdot (2\cdot b-1)}.\eqno{(12)}$$
Dividing both sides by $\sqrt{a\cdot (1-b)}$, we get an equivalent inequality
$$\frac{1}{\sqrt{(1-a)\cdot b}}\le \frac{4\cdot \sqrt{(1-a)\cdot b}}{(1-2\cdot a)\cdot (2\cdot b-1)}.\eqno{(13)}$$
Multiplying both sides by both denominators, we get the following equivalent inequality:
$$(1-2\cdot a)\cdot(2\cdot b-1)\le 4\cdot (1-a)\cdot b.\eqno{(14)}$$
If we open parentheses, this inequality takes the equivalent form
$$2\cdot b-4\cdot a\cdot b-1+2\cdot a\le 4\cdot b-4\cdot a\cdot b,\eqno{(15)}$$
i.e., by adding $4\cdot a\cdot b-2\cdot b$ to both sides, the form
$$-1+2\cdot a\le 2\cdot b.\eqno{(16)}$$ We are considering the case when $a\le b$ -- since, as we have mentioned earlier, it is sufficient to only consider this case. Thus, the equivalent inequality (12) is also true and hence, for the case when $\rho>0$, we indeed have $d\ge 0$.
\medskip

\noindent $4^\circ$. To complete the proof, it is now sufficient to consider the case when $\rho<0$. 
\medskip

In this case, if one of the values $a$ and $b$ is smaller than 0.5 and another one is larger than 0.5, then the differences $1-2\cdot a$ and $1-2\cdot b$ have different signs, so the right-hand side of the expression (5) for $d$ is larger than 1 and thus, non-negative. Thus, it is sufficient to consider the cases when:
\begin{itemize}
\item either both $a$ and $b$ are larger than 0.5 
\item or both $a$ and $b$ are smaller than 0.5. 
\end{itemize}
Let us consider these two cases one by one.
\medskip

\noindent $4.1^\circ$. Let us first consider the case when $a>0.5$ and $b>0.5$. 
\medskip

In this case, $a+b-1>0$, so the inequality (4) takes the form
$$a\cdot b+\rho\cdot \sqrt{a\cdot (1-a)\cdot b\cdot (1-b)}\ge a+b-1,\eqno{(17)}$$
i.e., equivalently, that
$$|\rho|\cdot  \sqrt{a\cdot (1-a)\cdot b\cdot (1-b)}\le a\cdot b-a-b+1=(1-a)\cdot (1-b),\eqno{(18)}$$
or that
$$|\rho|\le\frac{(1-a)\cdot (1-b)}{\sqrt{a\cdot (1-a)\cdot b\cdot (1-b)}}=\frac{\sqrt{(1-a)\cdot(1-b)}}{\sqrt{a\cdot b}}.\eqno{(19)}$$
In this case, the condition $d\ge 0$ that the value (5) is non-negative takes the form
$$1-|\rho|\cdot \frac{(2\cdot a-1)\cdot (2\cdot b-1)}{4\cdot \sqrt{a\cdot (1-a)\cdot b\cdot (1-b)}},\eqno{(20)}$$
i.e., equivalently,
$$|\rho|\cdot \frac{(2\cdot a-1)\cdot (2\cdot b-1)}{4\cdot \sqrt{a\cdot (1-a)\cdot b\cdot (1-b)}}\le 1\eqno{(21)}$$
and
$$|\rho|\le \frac{4\cdot \sqrt{a\cdot (1-a)\cdot b\cdot (1-b)}}{(2\cdot a-1)\cdot (2\cdot b-1)}.\eqno{(22)}$$
Similarly to the case when $\rho>0$, to check that all values $|\rho|$ satisfying the inequality (19) also satisfies the inequality (22), it is sufficient to check that the largest possible value $|\rho|$ satisfying the inequality (19) satisfies the inequality (22), i.e., that
$$\frac{\sqrt{(1-a)\cdot(1-b)}}{\sqrt{a\cdot b}}\le
\frac{4\cdot \sqrt{a\cdot (1-a)\cdot b\cdot (1-b)}}
{(2\cdot a-1)\cdot (2\cdot b-1)}.\eqno{(23)}$$
If we divide both sides by $\sqrt{(1-a)\cdot (1-b)}$, we get the following equivalent inequality
$$\frac{1}{\sqrt{a\cdot b}}\le \frac{4\cdot \sqrt{a\cdot b}}{(2\cdot a-1)\cdot (2\cdot b-1)}.\eqno{(24)}$$
Multiplying both sides by both denominators, we get the following equivalent inequality
$$(2\cdot a-1)\cdot (2\cdot b-1)\le 4\cdot a\cdot b.\eqno{(25)}$$
Opening parentheses, we get
$$4\cdot a\cdot b-2\cdot a-2\cdot b+1\le 4\cdot a\cdot b.\eqno{(26)}$$
Adding $2\cdot a+2\cdot b-4\cdot a\cdot b$ to both sides, we get an equivalent inequality
$$1\le 2\cdot a+2\cdot b,\eqno{(27)}$$ which is true since we consider the case when $a+b>1$.
So, in this case, we indeed have $d\ge 0$.
\medskip

\noindent $4.2^\circ$. Let us now consider the case when $a<0.5$ and $b<0.5$. 
\medskip

In this case, $a+b-1<0$, so the inequality (4) takes the form
$$a\cdot b+\rho\cdot \sqrt{a\cdot (1-a)\cdot b\cdot (1-b)}\ge 0,\eqno{(28)}$$
i.e., equivalently, that
$$|\rho|\cdot  \sqrt{a\cdot (1-a)\cdot b\cdot (1-b)}\le a\cdot b,\eqno{(29)}$$
or that
$$|\rho|\le\frac{a\cdot b}{\sqrt{a\cdot (1-a)\cdot b\cdot (1-b)}}=\frac{\sqrt{a\cdot b}}{\sqrt{(1-a)\cdot(1-b)}}.\eqno{(30)}$$
In this case, the condition $d\ge 0$ that the value (5) is non-negative takes the form
$$1-|\rho|\cdot \frac{(1-2\cdot a)\cdot (1-2\cdot b)}{4\cdot \sqrt{a\cdot (1-a)\cdot b\cdot (1-b)}},\eqno{(31)}$$
i.e., equivalently,
$$|\rho|\cdot \frac{(1-2\cdot a)\cdot (1-2\cdot b)}{4\cdot \sqrt{a\cdot (1-a)\cdot b\cdot (1-b)}}\le 1\eqno{(32)}$$
and
$$|\rho|\le \frac{4\cdot \sqrt{a\cdot (1-a)\cdot b\cdot (1-b)}}{(1-2\cdot a)\cdot (1-2\cdot b)}.\eqno{(33)}$$
Similarly to the cases when $\rho>0$ and when $a+b>1$, to check that all values $|\rho|$ satisfying the inequality (30) also satisfies the inequality (33), it is sufficient to check that the largest possible value $|\rho|$ satisfying the inequality (30) satisfies the inequality (33), i.e., that
$$\frac{\sqrt{a\cdot b}}{\sqrt{(1-a)\cdot(1-b)}}\le \frac{4\cdot \sqrt{a\cdot (1-a)\cdot b\cdot (1-b)}}{(1-2\cdot a)\cdot (1-2\cdot b)}.\eqno{(34)}$$
If we divide both sides by $\sqrt{a\cdot b}$, we get the following equivalent inequality
$$\frac{1}{\sqrt{(1-a)\cdot (1-b)}}\le \frac{4\cdot \sqrt{(1-a)\cdot (1-b)}}{(1-2\cdot a)\cdot (1-2\cdot b)}.\eqno{(35)}$$
Multiplying both sides by both denominators, we get the following equivalent inequality
$$(1-2\cdot a)\cdot (1-2\cdot b)\le 4\cdot (1-a)\cdot (1-b).\eqno{(36)}$$
Opening parentheses, we get
$$1-2\cdot a-2\cdot b+4\cdot a\cdot b\le 4-4\cdot a-4\cdot b+4\cdot a\cdot b.\eqno{(37)}$$
Adding $4\cdot a+4\cdot b-4\cdot a\cdot b-1$ to both sides, we get an equivalent inequality
$$2\cdot a+2\cdot b\le 3,\eqno{(38)}$$ which is true since we consider the case when $a+b<1$.
So, in this case, we indeed have $d\ge 0$.
\medskip

In all cases when have $d\ge 0$, thus, the and-operation $f_\rho(a,b)$ is indeed a copula. Thus, for boxes in which all four vertices belong to the area described by the expression (1), the inequality (4a) is always satisfied.
\medskip

\noindent $5^\circ$. Let us now consider the boxes in which two vertices belong to the boundary between two areas. First, we will consider the case when $\rho>0$ and then, we will consider the case when $\rho<0$. 
\medskip

\noindent $6^\circ$. Let us first consider the case when $\rho>0$. For this case, let us first describe the boundaries between the areas. 
\medskip

\noindent $6.1^\circ$. Let us analyze which of the three areas listed in formula (4) are possible in this case.
\medskip

When $\rho>0$, we have $$a\cdot b+\rho\cdot \sqrt{a\cdot (1-a)\cdot b\cdot (1-b)}\ge a\cdot b,$$ and since it is known that we always have $a\cdot b\ge \max(a+b-1,0)$, we have $$a\cdot b+\rho\cdot \sqrt{a\cdot (1-a)\cdot b\cdot (1-b)}\ge \max(a+b-1,0).$$ So, for $\rho>0$, we cannot have the first of the three cases described by the formula (4). So, we only have two areas:
\begin{itemize}
\item the area where the and-operation is described by the formula (1), and
\item the area where the and-operation is described by the formula $\min(a,b)$.
\end{itemize}
\medskip

\noindent $6.2^\circ$. 
Let us describe the two possible areas and the boundary between these two areas. 
\medskip

The first area is characterized by the inequality 
$$a\cdot b+\rho\cdot \sqrt{a\cdot (1-a)\cdot b\cdot (1-b)}\le \min(a,b).\eqno{(39)}$$
Similarly to the previous part of the proof, without losing generality, we can consider the case when $a\le b$. In this case, the inequality (39) describing the first area takes the following form:
$$a\cdot b+\rho\cdot \sqrt{a\cdot (1-a)\cdot b\cdot (1-b)}\le a.\eqno{(40)}$$
If we subtract $a\cdot b$ from both sides of this inequality, we get the following equivalent inequality:
$$\rho\cdot \sqrt{a\cdot (1-a)\cdot b\cdot (1-b)}\le a\cdot (1-b).\eqno{(41)}$$
Both sides of this inequality are non-negative, so we can get an equivalent inequality is we square both sides:
$$\rho^2\cdot a\cdot (1-a)\cdot b\cdot (1-b)\le a^2\cdot (1-b)^2.\eqno{(42)}$$
The cases when $a$ or $b$ are equal to 0 or 1 can be obtained by taking a limit from the cases when both $a$ and $b$ are located insyed the interval $(0,1)$. For such values, we can divide both side of the inequality by positive numbers $a^2$, $b$, and $1-b$, and get the following equivalent inequality:
$$\rho^2\cdot \frac{1-a}{a}\le\frac{1-b}{b},\eqno{(43)}$$
i.e., equivalently, 
$$\rho^2\cdot \frac{1-a}{a}\le\frac{1}{b}-1.\eqno{(44)}$$
By adding 1 to both sides of this inequality, we get
$$\frac{a+\rho^2\cdot (1-a)}{a}\le \frac{1}{b},\eqno{(45)}$$
i.e., equivalently, that 
$$b\le \frac{a}{a+\rho^2\cdot (1-a)}.\eqno{(46)}$$
This inequality describes the first area, in which the and-operation is described by the formula (1).
Thus, the boundary between the two areas is described by the equality 
$$b=\frac{a}{a+\rho^2\cdot (1-a)}.\eqno{(47)}$$
\medskip

\noindent{\it Comment.} 
One can see that for $a=0$ we get $b=0$, for $a=1$, we get $b=1$. 
\medskip

\noindent $6.3^\circ$. Let us prove that for all $a$, the corresponding boundary value $b$ is greater than or equal to $a$ -- i.e., that for all the points $(a,b)$ on this boundary, we have $a\le b$. 
\medskip

Indeed, for the expression (47), the desired inequality $a\le b$ takes the form
$$a\le \frac{a}{a+\rho^2\cdot (1-a)}.\eqno{(48)}$$
If we divide both sides by $a$ and multiply both sides by the denominator of the right-hand side, we get the following equivalent inequality
$$a+\rho^2\cdot (1-a)\le 1.\eqno{(49)}$$
If we move all the terms to the right-hand side, we get an equivalent inequality
$$0\le 1-a-\rho^2\cdot (1-a)=(1-\rho^2)\cdot (1-a).\eqno{(50)}$$
This inequality is always true, since $\rho^2\le 1$ and $a\le 1$, so indeed, for all boundary points, we have $a\le b$. 
\medskip

\noindent $6.4^\circ$. Let us prove that the boundary describes $b$ as an increasing function of $a$. 
\medskip

By applying, to the equality (47) that describes the boundary, the same transformations that show the equivalent of inequalities (43) and (46), we can conclude that the equality (47) is equivalent to
$$\rho^2\cdot \frac{1-a}{a}=\frac{1-b}{b},\eqno{(51)}$$
i.e., to
$$\rho^2\cdot \left(\frac{1}{a}-1\right)=\frac{1}{b}-1.\eqno{(52)}$$
The left-hand side is decreasing with respect to $a$, the right-hand side is a decreasing function of $b$. Thus, as $a$ increases, the left-hand side decreases, thus the right-hand side also decreases and hence, the value $b$ increases as well.
\medskip

\noindent $6.5^\circ$. For $\rho=1$ the condition (46) describing the first area takes the form $b\le a$. Since we have $a\le b$, this means that this condition is only satisfies for $a=b$. For these values, the expression (4a) is equal to $$a\cdot a+\sqrt{a\cdot(1-a)\cdot a\cdot (1-a)}=a^2+a\cdot (1-a)=a^2+a-a^2=$$ $$a=\min(a,b),\eqno{(53)}$$ which means that our and-operation is always equal to $\min(a,b)$. The expression $\min(a,b)$ is known to be a copula. 

So, we only need to prove the fact that our and-operation is a copula for the case when $\rho<1$. This is the case we will consider from now on.
\medskip

\noindent $6.6^\circ$. Let us prove that for $\rho<1$, the only boundary points for which $a=b$ are points for which $a=b=0$ and $a=b=1$.
\medskip

Indeed, as we have mentioned, the points $(0,0)$ and $(1,1)$ are boundary points. Let us prove, by contradiction, that there are no other boundary points for which $a=b$. Indeed, when $a=b$, the equality (52) that describes the boundary takes the form:
$$\rho^2\cdot \left(\frac{1}{a}-1\right)=\frac{1}{a}-1.\eqno{(54)}$$
Dividing both sides of this equality by the non-zero right-hand side, we get $\rho^2=1$. This contradicts to the fact that we are considering the case when $\rho<1$ and thus, $\rho^2<1$. This contradiction shows that other boundary points with $a=b$ are not possible. 
\medskip

\noindent $6.7^\circ$. The boundary consists of a curved line that is separate from the line $a=b$ -- except for the endpoints. So, if we limit ourselves to a sub-box $[\varepsilon,1-\varepsilon]\times [\varepsilon,1-\varepsilon]$ for some small $\varepsilon>0$, the boundary line is separated from the line $a=b$ -- there is the smallest distance $\delta>0$ between points of these two lines. So, if we have a box that includes both points with $a\le b$ and with $a\ge b$, we can divide this box into sub-boxes of linear size $<\delta/2$ and thus, make sure that every sub-box that contains boundary points with $a\le b$ cannot contain any points with $a=b$ -- and therefore, only contains points with $a\le b$. 

So, due to additivity, it is sufficient to prove the inequality (4a) for boxes for which:
\begin{itemize}
\item two vertices lie on the boundary, and 
\item we have $a\le b$ for all the points from this sub-box.
\end{itemize}
This will allow us to prove the inequality (4a) for all sub-boxes of the square $[\varepsilon,1-\varepsilon]\times [\varepsilon,1-\varepsilon]$. We can do it for any $\varepsilon$ and thus, in the limit, get the desired inequality for all sub-boxes of the original square $[0,1]\times [0,1]$ as well.

So, suppose that we have a box for which:
\begin{itemize}
\item two vertices lie on the boundary, and
\item we have $a\le b$ for all the points from this box.
\end{itemize}

Since the boundary describes the increasing function of $a$, the corresponding box has the form
\medskip

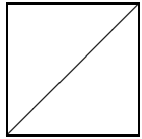
\begin{center}
\begin{picture}(50,50)
\put(0,0){\line(1,0){50}}
\put(0,50){\line(1,0){50}}
\put(0,0){\line(0,1){50}}
\put(50,0){\line(0,1){50}}
\put(0,0){\line(1,1){50}}
\end{picture}
\end{center}
\medskip

So, in the corresponding box:
\begin{itemize}
\item the two vertices $(\underline a,\underline b)$ and $(\overline a,\overline b)$ are on the boundary, 
\item the vertex $(\overline a,\underline b)$ is in the first area, i.e., for this point, we have the expression (1), and 
\item the vertex $(\underline a,\overline b)$ is in the second area, i.e., here $C(\underline a,\overline b)=\min(\underline a,\overline b)$. 
\end{itemize}

The desired inequality (4a) has the form 
$$C(\overline a,\underline b)-C(\underline a,\underline b)\le 
C(\overline a,\overline b)-C(\underline a,\overline b).\eqno{(55)}$$
The points $(\overline a,\overline b)$ and $(\underline a,\overline b)$ are both in the second area for which $C(a,b)=\min(a,b)$ -- to be more precise, the second of these points is in the boundary, which means it also satisfies the condition $C(a,b)=\min(a,b)$. For all the points from the box, $a\le b$, so we have
$$C(\overline a,\overline b)-C(\underline a,\overline b)=\min(\overline a,\overline b)-\min(\underline a,\overline b)=\overline a-\underline a.\eqno{(56)}$$
On the other hand, for the difference in the left-hand side of the formula (55), we have 
$$C(\overline a,\underline b)-C(\underline a,\underline b)=\int_{\underline a}^{\overline a} \frac{\partial C}{\partial a}\,da.\eqno{(57)}$$
So, if we prove that the partial derivative $\partial C/\partial a$ is always smaller or equal than 1, we would indeed conclude that 
$$C(\overline a,\underline b)-C(\underline a,\underline b)=\int_{\underline a}^{\overline a} 1\,da=\overline a-\underline a,\eqno{(58)}$$
i.e., exactly, the desired inequality (55). 

For the points $(\overline a,\underline b)$ and $(\underline a,\underline b)$ -- and the points from the interval connecting these two points -- the expression $C(a,b)$ is described by the formula (1). Thus, the partial derivative of $C(a,b)$ with respect to $a$ is described by the formula (4b). Thus, the inequality $$\frac{\partial C}{\partial a}(a,b)\le 1,\eqno{(59)}$$ takes the form 
$$b+\rho\cdot \frac{1-2\cdot a}{2\cdot\sqrt{a\cdot (1-a)}}\cdot \sqrt{b\cdot (1-b)}\le 1.\eqno{(60)}$$
Subtracting $b$ from both sides of (60), we get an equivalent inequality
$$\rho\cdot \frac{1-2\cdot a}{2\cdot\sqrt{a\cdot (1-a)}}\cdot \sqrt{b\cdot (1-b)}\le 1-b.\eqno{(61)}$$
To separate the variables, we can divide both sides by $\sqrt{b\cdot (1-b)}$, then we get an equivalent inequality
$$\rho\cdot \frac{1-2\cdot a}{2\cdot\sqrt{a\cdot (1-a)}}\le \sqrt{\frac{1-b}{b}}.\eqno{(62)}$$
By taking the square root of both sides of the inequality (46), we conclude that:
$$\rho\cdot \sqrt{\frac{1-a}{a}}\le \sqrt{\frac{1-b}{b}}.\eqno{(63)}$$
Thus, if we prove that the left-hand side of the inequality (62) is smaller than or equal to the left-hand side of the inequality (63), i.e., that 
$$\rho\cdot \frac{1-2\cdot a}{2\cdot\sqrt{a\cdot (1-a)}}\le \rho\cdot \sqrt{\frac{1-a}{a}};\eqno{(64)}$$
this will prove the inequality (62) and thus, the desired upper bound (60) on the partial derivative. We can simplify the inequality (64) by dividing both sides by $\rho$ and multiplying both sides by $2\cdot \sqrt{a\cdot (1-a)}$. Then, we get an equivalent inequality
$$1-2\cdot a\le 2\cdot (1-a)=2-2\cdot a,\eqno{(65)}$$ which is equivalent to $1\le 2$ and is, thus, always true. Thus, (55) holds, so the inequality (4a) is true for all the boxes in which two vertices are located on the boundary.

This completes the proof of the Proposition for the case when $\rho>0$. 
\medskip

\noindent $7^\circ$. Let us now consider the case when $\rho<0$. 
For this case, let us first describe the boundaries between the areas.
\medskip

\noindent $7.1^\circ$. Let us analyze which of the three areas listed in formula (4) are possible in this case.
\medskip

When $\rho<0$, we have $$a\cdot b+\rho\cdot \sqrt{a\cdot (1-a)\cdot b\cdot (1-b)}\le a\cdot b,$$ and since it is known that we always have $a\cdot b\le \min(a,b)$, we have $$a\cdot b+\rho\cdot \sqrt{a\cdot (1-a)\cdot b\cdot (1-b)}\le\min(a,b).$$ So, for $\rho<0$, we cannot have the third of the three cases described by the formula (4). So, we only have two areas:
\begin{itemize}
\item the area where the and-operation is described by the formula (1), and
\item the area where the and-operation is described by the formula $$\max(a+b-1,0).$$
\end{itemize}
\medskip

\noindent $7.2^\circ$.
Let us describe the two possible areas and the boundary between these two areas.
\medskip

The first area is characterized by the inequality $C(a,b)\ge \max(a+b-1,0)$, i.e., equivalently, by two inequalities 
$$a\cdot b-|\rho|\cdot \sqrt{a\cdot (1-a)\cdot b\cdot (1-b)}\ge 0\eqno{(66)}$$
and
$$a\cdot b-|\rho|\cdot \sqrt{a\cdot (1-a)\cdot b\cdot (1-b)}\ge a+b-1.\eqno{(67)}$$
Let us consider these two inequalities one by one. 
\medskip

\noindent $7.2.1^\circ$. The inequality (66) is equivalent to:
$$a\cdot b\ge |\rho|\cdot \sqrt{a\cdot (1-a)\cdot b\cdot (1-b)}.\eqno{(68)}$$
To separate the variables, let us divide both sides of this inequality by $$a\cdot \sqrt{b\cdot (1-b)},$$ then we get an equivalent inequality
$$\sqrt{\frac{b}{1-b}}\ge |\rho|\cdot \sqrt{\frac{1-a}{a}}.\eqno{(69)}$$
Both sides of this inequality are non-negative, thus if we square both sides, we get an equivalent inequality
$$\frac{b}{1-b}\ge \rho^2\cdot \frac{1-a}{a}.\eqno{(70)}$$
Reversing both sides, we get an equivalent inequality
$$\frac{1-b}{b}\le \frac{a}{\rho^2\cdot (1-a)},\eqno{(71)}$$
i.e., equivalently, 
$$\frac{1}{b}-1\le \frac{a}{\rho^2\cdot (1-a)}.\eqno{(72)}$$
By adding 1 to both sides, we get
$$\frac{1}{b}\le \frac{\rho^2\cdot (1-a)+a}{\rho^2\cdot (1-a)},\eqno{(73)}$$
i.e., equivalently, 
$$b\ge \frac{\rho^2\cdot (1-a)}{\rho^2\cdot (1-a)+a}.\eqno{(74)}$$
\medskip

\noindent $7.2.2^\circ$. The inequality (67) is equivalent to 
$$a\cdot b-a-b+1\ge |\rho|\cdot \sqrt{a\cdot (1-a)\cdot b\cdot (1-b)},\eqno{(75)}$$
i.e., 
$$(1-a)\cdot (1-b)\ge |\rho|\cdot \sqrt{a\cdot (1-a)\cdot b\cdot (1-b)}.\eqno{(76)}$$
To separate the variables, let us divide both sides by $(1-a)\cdot \sqrt{b\cdot (1-b)}$, then we get an equivalent inequality
$$\sqrt{\frac{1-b}{b}}\ge |\rho|\cdot \sqrt{\frac{a}{1-a}}.\eqno{(77)}$$
Both sides of this inequality are non-negative, thus if we square both sides, we get an equivalent inequality
$$\frac{1-b}{b}\ge \rho^2\cdot \frac{a}{1-a},\eqno{(78)}$$
i.e., equivalently, 
$$\frac{1}{b}-1\ge \rho^2\cdot \frac{a}{1-a}.\eqno{(79)}$$
By adding 1 to both sides, we get
$$\frac{1}{b}\ge \frac{\rho^2\cdot a + (1-a)}{1-a},\eqno{(80)}$$
i.e., equivalently,
$$b\le \frac{1-a}{\rho^2\cdot a+(1-a)}.\eqno{(81)}$$
\medskip

\noindent $7.2.3^\circ$. By combining the inequalities (74) and (81), we get the following description of the area in which the and-operation is described by the formula (1):
$$\frac{\rho^2\cdot (1-a)}{\rho^2\cdot (1-a)+a}\le b\le \frac{1-a}{\rho^2\cdot a+(1-a)}.\eqno{(82)}$$
Thus, the boundary between the two areas consists of the following two curves:
$$b=\frac{\rho^2\cdot (1-a)}{\rho^2\cdot (1-a)+a}\eqno{(83)}$$
and
$$b=\frac{1-a}{\rho^2\cdot a+(1-a)}.\eqno{(84)}$$
\medskip

\noindent $7.3^\circ$. Let us prove that:
\begin{itemize}
\item the curve (83) lies in the area where $a+b\le 1$, and 
\item the curve (84) lies in the area where $a+b\ge 1$.
\end{itemize}
\medskip

\noindent $7.3.1^\circ$. Let us first prove that for each value $b$ described by the formula (83), we have $a+b\le 1$. 
\medskip

We need to prove the inequality 
$$a+\frac{\rho^2\cdot (1-a)}{\rho^2\cdot (1-a)+a}\le 1.\eqno{(85)}$$
Subtracting $a$ from both sides, we get an equivalent inequality
$$\frac{\rho^2\cdot (1-a)}{\rho^2\cdot (1-a)+a}\le 1-a.\eqno{(86)}$$
Dividing both sides by $1-a$ and multiplying both sides by the denominator of the left-hand side, we get the following equivalent inequality:
$$\rho^2\le \rho^2\cdot (1-a)+a=\rho^2+(1-\rho^2)\cdot a,\eqno{(87)}$$ which is, of course, always true, since $\rho^2\le 1$ and $a\ge 0$. The statement is proven.
\medskip

\noindent $7.3.2^\circ$. Let us now prove that for each value $b$ described by the formula (84), we have $a+b\ge 1$.
\medskip

We need to prove the inequality
$$a+\frac{1-a}{\rho^2\cdot a+(1-a)}\ge 1.\eqno{(88)}$$
Subtracting $a$ from both sides, we get an equivalent inequality
$$\frac{1-a}{\rho^2\cdot a+(1-a)}\ge 1-a.\eqno{(89)}$$
Dividing both sides by $1-a$ and multiplying both sides by the denominator of the left-hand side, we get the following equivalent inequality:
$$1 \ge \rho^2\le a+(1-a)=1-(1-\rho^2)\cdot a,\eqno{(90)}$$ which is, of course, always true. The statement is proven. 
\medskip

\noindent $7.4^\circ$. Similarly to Part 6 of this proof, it is sufficient to prove the inequality (4a) for boxes in which two vertices are on the boundary and for which:
\begin{itemize}
\item either we have $a+b\le 1$ for all the points from the box, 
\item or we have $a+b\ge 1$ for all the points from the box.
\end{itemize}
Let us consider the two parts of the boundary one by one.
\medskip

\noindent $7.4.1^\circ$. Let us first consider the case when we have $a+b\le 1$ for all the points from the box. In this case, the corresponding part of the boundary is described by the formula (83). By reformulating this expression in the equivalent form
$$b=\frac{1}{1+\displaystyle\frac{a}{\rho^2\cdot (1-a)}}=
\frac{1}
{
1+\displaystyle\frac{1}{\rho^2}\cdot 
  \displaystyle\frac{1}{\displaystyle\frac{1}{a}-1}
},\eqno{(91)}$$
we can see that $b$ is a decreasing function of $a$. Thus, the corresponding box has the form
\medskip

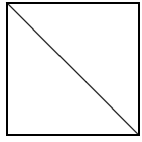
\begin{center}
\begin{picture}(50,50)
\put(0,0){\line(1,0){50}}
\put(0,50){\line(1,0){50}}
\put(0,0){\line(0,1){50}}
\put(50,0){\line(0,1){50}}
\put(0,50){\line(1,-1){50}}
\end{picture}
\end{center}
\medskip

So, in the corresponding box:
\begin{itemize}
\item the two vertices $(\underline a,\overline b)$ and $(\overline a,\underline b)$ are on the boundary,
\item the vertex $(\overline a,\overline b)$ is in the first area, i.e., for this point, we have the expression (1), and
\item the vertex $(\underline a,\underline b)$ is in the second area, i.e., here $C(\underline a,\overline b)=\max(\underline a+\underline b-1,0)$.
\end{itemize}
So, for three vertices, we have $C(a,b)=\max(a+b-1,0)$. Since for all the points from the box, we have $a+b\le 1$, this means that for three vertices, we have $C(a,b)=0$. In this case, the inequality (4a) is clearly true. 
\medskip

\noindent $7.4.2^\circ$. Let us now consider the case when we have $a+b\ge 1$ for all the points from the box. In this case, the corresponding part of the boundary is described by the formula (84). By reformulating this expression in the equivalent form
$$b=\frac{1}{1+\rho^2\cdot \displaystyle\frac{a}{1-a}}=
\frac{1}
{
1+\rho^2\cdot
  \displaystyle\frac{1}{\displaystyle\frac{1}{a}-1}
},\eqno{(91)}$$
we can see that $b$ is also a decreasing function of $a$. Thus, the corresponding box has the same form as in the case $a+b\le 1$: 
\medskip

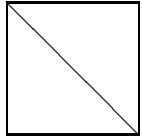
\begin{center}
\begin{picture}(50,50)
\put(0,0){\line(1,0){50}}
\put(0,50){\line(1,0){50}}
\put(0,0){\line(0,1){50}}
\put(50,0){\line(0,1){50}}
\put(0,50){\line(1,-1){50}}
\end{picture}
\end{center}
\medskip

So, in the corresponding box:
\begin{itemize}
\item the two vertices $(\underline a,\overline b)$ and $(\overline a,\underline b)$ are on the boundary,
\item the vertex $(\underline a,\underline b)$ is in the first area, i.e., for this point, we have the expression (1), and
\item the vertex $(\overline a,\overline b)$ is in the second area, i.e., here $C(\underline a,\overline b)=\max(\underline a+\underline b-1,0)$.
\end{itemize}
Similarly to Part 6 of the proof, we can show that the desired inequality (4a) is satisfied if we the corresponding partial derivatives is smaller than or equal to 1, i.e., if 
$$\frac{\partial C}{\partial a}=b-|\rho|\cdot \frac{1-2\cdot a}{2\cdot \sqrt{a\cdot (1-a)}}\cdot \sqrt{b\cdot (1-b)}\le 1.\eqno{(92)}$$
Subtracting $b$ from both sides, we get an equivalent inequality 
$$-|\rho|\cdot \frac{1-2\cdot a}{2\cdot \sqrt{a\cdot (1-a)}}\cdot \sqrt{b\cdot (1-b)}\le 1-b.\eqno{(93)}$$
We can separate the variable if we divide both sides by $\sqrt{b\cdot (1-b)}$, then we get an equivalent inequality 
$$-|\rho|\cdot \frac{1-2\cdot a}{2\cdot \sqrt{a\cdot (1-a)}}\le \sqrt{\frac{1-b}{b}}.\eqno{(94)}$$
We know a lower bound on the expression in the right-hand side -- it is provided by the inequality (77). Thus, to prove the inequality (94), it is sufficient to prove that the left-hand side of the formula (94) is smaller than or equal to this lower bound, i.e., that
$$-|\rho|\cdot \frac{1-2\cdot a}{2\cdot \sqrt{a\cdot (1-a)}}
\le|\rho|\cdot \sqrt{\frac{a}{1-a}}.\eqno{(95)}$$
Let us prove this inequality. Dividing both sides of (95) by $|\rho|$ and multiplying both sides by $2\cdot \sqrt{a\cdot (1-a)}$, we get an equivalent inequality $-(1-2\cdot a)\le 2\cdot a$, i.e., $2\cdot a-1\le 2\cdot a$, which is always true. Thus, the inequality (94) holds, hence the inequality (92) also holds, and therefore, in this case, the inequality (4a) that describes a copula is also true.
\medskip

\noindent $8^\circ$. We have considered all possible cases, and in all these cases, we have shown that the inequality (4a) -- that defines a copula -- is true. Thus, our and-operation is indeed a copula. The proposition is proven.

\section*{Acknowledgments}

This research was partly funded by the
EPSRC and ESRC CDT in Risk and Uncertainty (EP/L015927/1), established within
the Institute for Risk and Uncertainty at the University of Liverpool. This work has
been carried out within the framework of the EUROfusion Consortium, funded by the
European Union via the Euratom Research and Training Programme (Grant Agreement
No 101052200 - EUROfusion).

Views and opinions expressed are however those
of the author(s) only and do not necessarily reflect those of the European Union or the
European Commission. Neither the European Union nor the European Commission
can be held responsible for them.

V.K. was supported in part by the National Science Foundation
grants 1623190 (A Model of Change for Preparing a New Generation
for Professional Practice in Computer Science), and HRD-1834620 and
HRD-2034030 (CAHSI Includes), and by the AT\&T Fellowship in
Information Technology. He was also supported by the program of the
development of the Scientific-Educational Mathematical Center of
Volga Federal District No. 075-02-2020-1478, and by a grant from
the Hungarian National Research, Development and Innovation Office
(NRDI).

\end{document}